# EXAMPLES OF MODELS OF THE ASYNCHRONOUS CIRCUITS

Serban E. VLAD
Oradea City Hall, P-ta Unirii, Nr. 1, 410100, Oradea
serban_e_vlad@yahoo.com, www.oradea.ro/serban/index.html

**Abstract.**
The notion of limit condition that we propose is a generalization of the delay condition [1], [2], including the models of the delay circuits, of the C elements of Muller and of the combinational circuits.

**1. Preliminaries**

1.1 **Definition** $B = \{0,1\}$ is endowed with the order $0 \leq 1$ and with the usual laws: $\overline{\phantom{x}}, \cdot, \cup, \oplus$. They induce an order and laws on the set of the $R \to B$ functions that are noted with the same symbols.

1.2 **Definition** Let $x: R \to B$ and $A \subset R$, $A \neq \varnothing$. We define
$$\bigcap_{\xi \in A} x(\xi) = \begin{cases} 0, \exists \xi \in A, x(\xi) = 0 \\ 1, otherwise \end{cases}, \quad \bigcup_{\xi \in A} x(\xi) = \begin{cases} 1, \exists \xi \in A, x(\xi) = 1 \\ 0, otherwise \end{cases}$$

1.3 **Definition** Let $x: R \to B$. The *left limit* function $x(t-0)$ is defined by:
$$\forall t \in R, \exists \varepsilon > 0, \forall \xi \in (t - \varepsilon, t), x(\xi) = x(t-0)$$

1.4 **Definition** The functions $\overline{x(t-0)} \cdot x(t), \; x(t-0) \cdot \overline{x(t)}$ are called the *left semi-derivatives* of $x$.

1.5 **Definition** The *characteristic function* $\chi_A : R \to B$ of the set $A \subset R$ is defined by:
$$\chi_A(t) = \begin{cases} 1, t \in A \\ 0, t \notin A \end{cases}$$

1.6 **Definition** We call *signal* a function $x$ having the property that the unbounded sequence $0 \leq t_0 < t_1 < t_2 < ...$ exists so that
$$x(t) = x(t_0 - 1) \cdot \chi_{(-\infty, t_0)}(t) \oplus x(t_0) \cdot \chi_{[t_0, t_1)}(t) \oplus x(t_1) \cdot \chi_{[t_1, t_2)}(t) \oplus ...$$

We note with $S^{(m)}, m \geq 1$ the set of the $R \to B^m$ functions whose coordinates are signals and sometimes we write $S$ instead of $S^{(1)}$.

1.7 **Theorem** $\forall x \in S,$ the left limit function $x(t-0)$ exists.

**1.8 Notation** We note with $\tau^d : R \to R$ the translation $\tau^d(t) = t - d$, where $t, d \in R$.

**1.9 Theorem** $\forall u \in S^{(m)}, \forall d \geq 0$ we have $u \circ \tau^d \in S^{(m)}$. Let $f : B^m \to B$ and the real numbers $0 \leq m' \leq d'$. For all $u$, the functions $f(u(t))$, $\bigcap_{\xi \in [t-d', t-d'+m']} f(u(\xi))$, $\bigcup_{\xi \in [t-d', t-d'+m']} f(u(\xi))$ are signals.

**1.10 Notation** We note with $P^*(S)$ the set of the non-empty subsets of $S$.

## 2. Limit Condition

**2.1 Definition** Let $f, g : B^m \to B$ two functions with the property $\forall a \in B^m, f(a) \leq g(a)$ and we identify the statement
$$\forall \lambda \in B, \forall \mu \in B, (\exists t_1, \forall t \geq t_1, f(u(t)) = \lambda \text{ and } g(u(t)) = \mu) \Rightarrow (\exists t_2, \forall t \geq t_2, \lambda \leq x(t) \leq \mu)$$
where $u \in S^{(m)}, x \in S$ with the function $Sol_{f,g} : S^{(m)} \to P^*(S)$ defined by
$$Sol_{f,g}(u) = \{x \mid \forall \lambda \in B, \forall \mu \in B,$$
$$(\exists t_1, \forall t \geq t_1, f(u(t)) = \lambda \text{ and } g(u(t)) = \mu) \Rightarrow (\exists t_2, \forall t \geq t_2, \lambda \leq x(t) \leq \mu)\}$$
A function $i : S^{(m)} \to P^*(S)$ with the property
$$\forall u, i(u) \subset Sol_{f,g}(u)$$
is called *limit condition* induced by $f, g$ ($LC_{f,g}$); $u$ is called *input* and $x \in i(u)$ is called *state* or *output*.

**2.2 Special cases** a) $m = 1$ and $f = g = 1_B$; the condition
$$\forall \lambda \in B, (\exists t_1, \forall t \geq t_1, u(t) = \lambda) \Rightarrow (\exists t_2, \forall t \geq t_2, x(t) = \lambda)$$
is called the *stability condition* (SC); $Sol_{1_B, 1_B}$ is noted $Sol_{SC}$ and the $LC_{1_B, 1_B}$'s are called *delay conditions* ($DC$'s), or shortly *delays*.

b) $f(a_1, ..., a_m) = a_1 \cdot ... \cdot a_m$, $g(a_1, ..., a_m) = a_1 \cup ... \cup a_m$. The condition
$$\forall \lambda \in B, \forall \mu \in B, (\exists t_1, \forall t \geq t_1, u_1(t) \cdot ... \cdot u_m(t) = \lambda \text{ and } u_1(t) \cup ... \cup u_m(t) = \mu) \Rightarrow$$
$$\Rightarrow (\exists t_2, \forall t \geq t_2, \lambda \leq x(t) \leq \mu)$$
is called the *Muller condition* (MC); $Sol_{f,g}$ is noted $Sol_{MC}$ and the $LC_{f,g}$'s are called $m - delay$ *conditions* ($m - DC$'s) or shortly $m - delays$.

c) $f = g$; the condition
$$\forall \lambda \in B, (\exists t_1, \forall t \geq t_1, f(u(t)) = \lambda) \Rightarrow (\exists t_2, \forall t \geq t_2, x(t) = \lambda)$$
is called the *stability condition* induced by $f$ ($SC_f$); $Sol_{f,f}$ is noted $Sol_{SC_f}$ and the $LC_{f,f}$'s are called $f - delay$ *conditions* ($f - DC$'s), or shortly $f - delays$.

**2.3 Remark** The $LC_{f,g}$'s model the asynchronous circuits, in the sense that if certain demands, as expressed by the existence of $\lambda, \mu$ act persistently on $x$, then $x$ obeys these demands persistently. The $DC$'s model the delay circuits, the $m - DC$'s model the C circuits of Muller and the $f - DC$'s model the combinational circuits that compute $f$.

Generally, the circuits are modeled by systems of equations and inequalities and in such a situation for any $u$, the set $i(u)$ consists in the solutions of the system.

**2.4 Theorem** a) $\forall u \in S^{(m)}, Sol_{SC_f}(u) = Sol_{SC}(f(u))$

b) $\forall v \in S, Sol_{MC}(v,...,v) = Sol_{SC}(v)$

$\forall u \in S^{(m)}, \forall p \in \{1,...,m\}, Sol_{SC}(u_p) \subset Sol_{MC}(u)$

c) We suppose that $f$ satisfies

$$\forall a \in B^m, a_1 \cdot ... \cdot a_m \leq f(a_1,...,a_m) \leq a_1 \cup ... \cup a_m$$

Then we have

$$\forall u \in S^{(m)}, Sol_{SC_f}(u) \subset Sol_{MC}(u)$$

## 3. Order, Determinism, Time Invariance, Constancy, Symmetry, Direct Product and Serial Connection

**3.1 Definition** We consider the functions $f, g, f', g': B^m \to B$, $m \geq 1$ with the property $\forall a \in B^m, f(a) \leq g(a), f'(a) \leq g'(a)$ the $LC_{f,g}$ $i: S^{(m)} \to P^*(S)$ and the $LC_{f',g'}$. $j: S^{(m)} \to P^*(S)$. We define the order $i \subset j$ by $\forall u, i(u) \subset j(u)$.

**3.2 Definition** The $LC_{f,g}$ $i$ is called:

a) *deterministic* if $\forall u, i(u)$ has a single element and *non-deterministic* otherwise.

b) *time invariant* if it fulfills

$$\forall u, \forall x, \forall d \in R, (u \circ \tau^d \in S^{(m)} \text{ and } x \in i(u)) \Rightarrow (x \circ \tau^d \in S \text{ and } x \circ \tau^d \in i(u \circ \tau^d))$$

and *time variable* otherwise

c) *constant* if $d_r \geq 0, d_f \geq 0$ exist so that $\forall u, \forall x \in i(u)$ we have

$$\overline{x(t-0)} \cdot x(t) \leq g(u(t-d_r))$$
$$x(t-0) \cdot \overline{x(t)} \leq \overline{f(u(t-d_f))}$$

and *non-constant* otherwise

d) *symmetrical* (in the usual sense) if for any bijection $\sigma: \{1,...,m\} \to \{1,...,m\}$ and any $u \in S^{(m)}$ the property

$$i(u_1,...,u_m) = i(u_{\sigma(1)},...,u_{\sigma(m)})$$

is fulfilled and *asymmetrical* (in the usual sense) otherwise

e) *symmetrical* (in the rising-falling sense) if

$$\forall u, \forall x, x \in i(u_1,...,u_m) \Leftrightarrow \overline{x} \in i(\overline{u_1},...,\overline{u_m})$$

and *asymmetrical* (in the rising-falling sense) otherwise.

**3.3 Remarks** By interpreting $i$ as the set of the solutions of a system, the inclusion $i \subset j$ means the fact that the first system contains more restrictive conditions than the second and the model in the first case is more precise than in the second one. In particular a deterministic $LC_{f,g}$ contains the maximal information and the $LC_{f,g}$ $Sol_{f,g}$ contains the minimal information about the modeled circuit. Any $LC_{f,g}$ $j$ includes a deterministic $LC_{f,g}$ $i$.

Determinism indicates the uniqueness of the solution for all $u$. On the other hand we can consider the deterministic $LC_{f,g}$'s be $S^{(m)} \to S$ functions. The non-deterministic $LC_{f,g}$'s are justified by the fact that in an electrical circuit to one input $u$ there correspond several possible outputs $x$ depending on the variations in ambient temperature, power supply, on the technology etc.

Constancy means that $x$ is allowed to switch only if $g(u), f(u)$ have anticipated this possibility $d_r$, respectively $d_f$ time units before. Its satisfaction does not imply the uniqueness of $d_r, d_f$.

**3.4 Theorem** Let $V \subset S$, the arbitrary function $\varphi: S^{(m)} \to P^*(S)$, $f, g: \boldsymbol{B}^m \to \boldsymbol{B}$ with $\forall a \in \boldsymbol{B}^m, f(a) \leq g(a)$ and the $LC_{f,g}$'s $i, j$.

  a) If $\forall u, i(u) \wedge V \neq \varnothing$, then the next equation defines an $LC_{f,g}$:
$$(i \wedge V)(u) = i(u) \wedge V$$
  b) If $i, j$ satisfy $\forall u, i(u) \wedge j(u) \neq \varnothing$, then $i \wedge j$ is an $LC_{f,g}$ defined by
$$(i \wedge j)(u) = i(u) \wedge j(u)$$
  c) If $\forall u, i(u) \wedge \varphi(u) \neq \varnothing$ is true, then $i \wedge \varphi$ is an $LC_{f,g}$:
$$(i \wedge \varphi)(u) = i(u) \wedge \varphi(u)$$
  d) $i$ and $j$ define the $LC_{f,g}$ $i \vee j$ in the next manner:
$$(i \vee j)(u) = i(u) \vee j(u)$$

**3.5 Theorem** If $i$ is a time invariant $LC_{f,g}$, then the next equivalence holds:
$$\forall u, \forall x, \forall d \geq 0, x \in i(u) \Leftrightarrow x \circ \tau^d \in i(u \circ \tau^d)$$

**3.6 Examples** Let $h: \boldsymbol{B}^m \to \boldsymbol{B}$ with the property $\forall a \in \boldsymbol{B}^m, f(a) \leq h(a) \leq g(a)$. The function $I_d^h: S^{(m)} \to P^*(S)$ defined for $d \geq 0$ by $I_d^h(u) = \{h(u \circ \tau^d)\}$ is a deterministic, time invariant, constant $LC_{f,g}$. It is symmetrical in the usual sense iff for all bijections $\sigma: \{1, ..., m\} \to \{1, ..., m\}$ we have $\forall a \in \boldsymbol{B}^m$, $h(a_1, ..., a_m) = h(a_{\sigma(1)}, ..., a_{\sigma(m)})$ and it is symmetrical in the rising-falling sense iff $\forall a \in \boldsymbol{B}^m$, $\overline{h(a_1, ..., a_m)} = h(\overline{a_1}, ..., \overline{a_m})$.

$Sol_{f,g}$ is a non-deterministic, time variable, non-constant $LC_{f,g}$. The symmetry is equivalent with: for all bijections $\sigma: \{1, ..., m\} \to \{1, ..., m\}$ we have $\forall a \in \boldsymbol{B}^m$, $f(a_1, ..., a_m) = f(a_{\sigma(1)}, ..., a_{\sigma(m)})$, $g(a_1, ..., a_m) = g(a_{\sigma(1)}, ..., a_{\sigma(m)})$ respectively with $\forall a \in \boldsymbol{B}^m$, $\overline{f(a_1, ..., a_m)} = g(\overline{a_1}, ..., \overline{a_m})$.

Let $V \subset S$ and the $LC_{f,g}$'s $i, j, k$. We suppose that $i$ is deterministic. If $\forall u, i(u) \wedge V \neq \varnothing$, then $i \wedge V (= i)$ is deterministic and if $\forall u, i(u) \wedge j(u) \neq \varnothing$, then $i \wedge j (= i)$ is deterministic.

We suppose that $i, j$ are time invariant with $\forall u, i(u) \wedge j(u) \neq \varnothing$; then $i \wedge j$ is time invariant. If $k$ is time invariant; then $i \vee k$ is time invariant.

We suppose that $i$ is constant. If $i \wedge V$ and $i \wedge j$ are defined, then they are constant. More general, any $LC_{f,g}$ included in a constant $LC_{f,g}$ is constant. We suppose that $j$ is constant also and that $\forall u, i(u) \wedge j(u) = \varnothing$; then $i \vee j$ is not constant, in general. But if $\exists u, i(u) \wedge j(u) \neq \varnothing$, then $i \vee j$ is constant.

We suppose that $i, j$ are symmetrical in the usual sense (in the rising-falling sense); if $i \wedge j$ is defined, then it is symmetrical in the usual sense (in the rising-falling sense). The $LC_{f,g}$ $i \vee j$ is symmetrical in the usual sense (in the rising-falling sense) too.

**3.7 Definition** The *direct product* of $f_p : \boldsymbol{B}^{n_p} \to \boldsymbol{B}, n_p \geq 1, p = \overline{1,m}$ is the function $(f_1,..., f_m) : \boldsymbol{B}^{n_1+...+n_m} \to \boldsymbol{B}^m$, $(f_1,..., f_m)(a^1,..., a^m) = (f_1(a^1),..., f_m(a^m))$.

**3.8 Definition** Let $f_p, g_p : \boldsymbol{B}^{n_p} \to \boldsymbol{B}, n_p \geq 1$ with $\forall a^p \in \boldsymbol{B}^{n_p}, f_p(a^p) \leq g_p(a^p)$ and the $LC_{f_p, g_p}$'s $j_p$, $p = \overline{1,m}$. The function $(j_1,..., j_m) : S^{(n_1+...+n_m)} \to P^*(S)^m$, $(j_1,..., j_m)(u^1,..., u^m) = (j_1(u^1),..., j_m(u^m))$ is called the *direct product* of $j_1,..., j_m$.

**3.9 Definition** Let the $LC_{f,g}$ $i$, the $LC_{f_p, g_p}$'s $j_p, p = \overline{1,m}$ and we suppose that $\forall (a^1,..., a^m) \in \boldsymbol{B}^{n_1+...+n_m}, f \circ (f_1,..., f_m)(a^1,..., a^m) \leq g \circ (g_1,..., g_m)(a^1,..., a^m)$. The *serial connection* of $i$ and $(j_1,..., j_m)$ is noted $k = i \circ (j_1,..., j_m)$ and is defined by $k : S^{(n_1+...+n_m)} \to P^*(S)$,

$k(u^1,..., u^m) = \{x \mid \exists y_1,..., \exists y_m, \ y_1 \in j_1(u^1) \text{ and } ... \text{ and } y_m \in j_m(u^m) \text{ and } x \in i(y_1,..., y_m)\}$

**3.10 Theorem** a) If $i, j_1$ are $DC$'s (if $i, j_1,..., j_m$ are $m - DC$, $n_1 - DC,..., n_m - DC$, respectively if $i, j_1,..., j_m$ are $f - DC, f_1 - DC,..., f_m - DC$), then $k$ is defined and it is a $DC$ (an $n_1 + ... + n_m - DC$, respectively an $f \circ (f_1,..., f_m) - DC$). If one of $f, g$ satisfy

$$\forall a \in \boldsymbol{B}^m, \forall a' \in \boldsymbol{B}^m, (a_1 \leq a_1' \text{ and } ... \text{ and } a_m \leq a_m') \Rightarrow f(a_1,..., a_m) \leq f(a_1',..., a_m') \quad (1)$$

then $k$ exists and it is a $LC_{f \circ (f_1,..., f_m), g \circ (g_1,..., g_m)}$.

b) If $i, j_1,..., j_m$ are deterministic, then $k$ is deterministic.

c) If $i, j_1,..., j_m$ are time invariant, then $k$ is time invariant.

d) If $i, j_1,..., j_m$ are symmetrical in the rising-falling sense, then $k$ is symmetrical in the rising-falling sense.

**3.11 Remark** At Theorem 3.10 if $i, j_1,..., j_m$ are constant (symmetrical in the usual sense), then $k$ is not necessarily constant (symmetrical in the usual sense).

**3.12 Theorem** Let $V, V_1,..., V_m \subset S$, the $LC_{f,g}$'s $i, j$ and the $LC_{f_p, g_p}$'s $k_p, l_p, p = \overline{1,m}$.

a) The next implications are true:
$$i \subset j \Rightarrow i \circ (k_1,..., k_m) \subset j \circ (k_1,..., k_m)$$
$$k_1 \subset l_1 \text{ and } ... \text{ and } k_m \subset l_m \Rightarrow i \circ (k_1,..., k_m) \subset i \circ (l_1,..., l_m)$$

b) If $\forall u, i(u) \wedge V \neq \emptyset$, then $\forall (u^1,..., u^m), (i \circ (k_1,..., k_m))(u^1,..., u^m) \wedge V \neq \emptyset$ and
$$(i \wedge V) \circ (k_1,..., k_m) = (i \circ (k_1,..., k_m)) \wedge V$$

If $\forall p, \forall u^p, k_p(u^p) \wedge V_p \neq \emptyset$, then by noting with $i_{|V_1 \times ... \times V_m}$ the restriction of $i$ at $V_1 \times ... \times V_m$, we have
$$i \circ (k_1 \wedge V_1,..., k_m \wedge V_m) = i_{|V_1 \times ... \times V_m} \circ (k_1,..., k_m)$$

c) If $\forall u, i(u) \wedge j(u) \neq \emptyset$, then
$\forall (u^1,..., u^m), (i \circ (k_1,..., k_m))(u^1,..., u^m) \wedge (j \circ (k_1,..., k_m))(u^1,..., u^m) \neq \emptyset$ and
$$(i \wedge j) \circ (k_1,..., k_m) \subset (i \circ (k_1,..., k_m)) \wedge (j \circ (k_1,..., k_m))$$

If $\forall p, \forall u^p, k_p(u^p) \wedge l_p(u^p) \neq \emptyset$, then

$$\forall (u^1,...,u^m), (i \circ (k_1,...,k_m))(u^1,...,u^m) \wedge (i \circ (l_1,...,l_m))(u^1,...,u^m) \neq \emptyset \text{ and}$$
$$i \circ (k_1 \wedge l_1,...,k_m \wedge l_m) \subset (i \circ (k_1,...,k_m)) \wedge (i \circ (l_1,...,l_m))$$

d)
$$(i \vee j) \circ (k_1,...,k_m) = (i \circ (k_1,...,k_m)) \vee (j \circ (k_1,...,k_m))$$
$$i \circ (k_1 \vee l_1,...,k_m \vee l_m) \supset (i \circ (k_1,...,k_m)) \vee (i \circ (l_1,...,l_m))$$

## 4. Boundness

**4.1 Theorem** Let $f,g : \boldsymbol{B}^m \to \boldsymbol{B}$ with $\forall a \in \boldsymbol{B}^m, f(a) \leq g(a)$ and the numbers $0 \leq m_r \leq d_r$, $0 \leq m_f \leq d_f$.

a) When $u$ runs in $S^{(m)}$, the next functions
$$x(t) = \bigcap_{\xi \in [t-d_r, t-d_r+m_r]} f(u(\xi)), \quad x(t) = \bigcup_{\xi \in [t-d_f, t-d_f+m_f]} g(u(\xi))$$
define deterministic, time invariant, constant $LC_{f,g}$'s.

b) The next system
$$\bigcap_{\xi \in [t-d_r, t-d_r+m_r]} f(u(\xi)) \leq x(t) \leq \bigcup_{\xi \in [t-d_f, t-d_f+m_f]} g(u(\xi)) \tag{1}$$
defines a $LC_{f,g}$ iff one of the next statements is true:
$$d_r - m_r \leq d_f, d_f - m_f \leq d_r \tag{2}$$
$$\max_{a \in \boldsymbol{B}^m} f(a) \leq \min_{a \in \boldsymbol{B}^m} g(a) \tag{3}$$

**4.2 Definition** The system 4.1 (1) with the hypothesis 4.1 (2) or 4.1 (3) fulfilled is called the *bounded limit condition* induced by $f,g$ ($BLC_{f,g}$). We identify it with the function $Sol_{BLC_{f,g}}^{m_r,d_r,m_f,d_f} : S^{(m)} \to P^*(S)$ defined by

$$Sol_{BLC_{f,g}}^{m_r,d_r,m_f,d_f}(u) = \{x \mid x \text{ satisfies } 4.1\,(1)\}$$

**4.3 Theorem** The next statements are equivalent:

a) $Sol_{BLC_{f,g}}^{m_r,d_r,m_f,d_f}$ is deterministic

b) Exactly one of b.1), b.2) takes place
    b.1) $f = g = c$ (the constant function)
    b.2) $f = g$ non-constant and one of the next equivalent properties is true
        b.2.1) $d_r = d_f - m_f, d_f = d_r - m_r$
        b.2.2) $m_r = m_f = 0$
        b.2.3) $\exists d \geq 0, Sol_{BLC_{f,g}}^{m_r,d_r,m_f,d_f} = I_d^f$

**4.4 Theorem** The next statements are equivalent

a) $Sol_{BLC_{f,g}}^{m_r,d_r,m_f,d_f} \subset Sol_{BLC_{f',g'}}^{m_r',d_r',m_f',d_f'}$

b) $\forall a \in \boldsymbol{B}^m, f'(a) \leq f(a) \leq g(a) \leq g'(a)$ and moreover we have

$$\max_{a \in B^m} f'(a) \leq \min_{a \in B^m} f(a) \text{ or } d'_r - m'_r \leq d_r - m_r \leq d_r \leq d'_r$$

$$\max_{a \in B^m} g(a) \leq \min_{a \in B^m} g'(a) \text{ or } d'_f - m'_f \leq d_f - m_f \leq d_f \leq d'_f$$

**4.5 Theorem** $Sol^{m_r,d_r,m_f,d_f}_{BLC_{f,g}}$ is time invariant

**4.6 Theorem** $Sol^{m_r,d_r,m_f,d_f}_{BLC_{f,g}}$ is symmetrical in the usual sense iff for all bijections

$\sigma: \{1,...,m\} \to \{1,...,m\}$ and all $a \in B^m$ we have
$$f(a_1,...,a_m) = f(a_{\sigma(1)},...,a_{\sigma(m)}), \ g(a_1,...,a_m) = g(a_{\sigma(1)},...,a_{\sigma(m)})$$

**4.7 Theorem** $Sol^{m_r,d_r,m_f,d_f}_{BLC_{f,g}}$ is symmetrical in the rising-falling sense iff $d_r = d_f, m_r = m_f$

and $\forall a \in B^m, \overline{f(a_1,...,a_m)} = g(\overline{a_1},...,\overline{a_m})$.

**4.8 Theorem** We consider $Sol^{m_r,d_r,m_f,d_f}_{BLC_{f,g}}, Sol^{m'_r,d'_r,m'_f,d'_f}_{BLC_{f_1,g_1}},..., Sol^{m'_r,d'_r,m'_f,d'_f}_{BLC_{f_m,g_m}}$ so that $f, g$

satisfy the additional monotony requests 3.10 (1) as well as

$$\forall u \in S^{(m)}, f(\bigcap_{\xi \in [t-d_r, t-d_r+m_r]} u_1(\xi),..., \bigcap_{\xi \in [t-d_r, t-d_r+m_r]} u_m(\xi)) = \bigcap_{\xi \in [t-d_r, t-d_r+m_r]} f(u_1(\xi),...,u_m(\xi))$$

$$\forall u \in S^{(m)}, g(\bigcup_{\xi \in [t-d_f, t-d_f+m_f]} u_1(\xi),..., \bigcup_{\xi \in [t-d_f, t-d_f+m_f]} u_m(\xi)) = \bigcup_{\xi \in [t-d_f, t-d_f+m_f]} g(u_1(\xi),...,u_m(\xi))$$

We have:
$$Sol^{m_r,d_r,m_f,d_f}_{BLC_{f,g}} \circ (Sol^{m'_r,d'_r,m'_f,d'_f}_{BLC_{f_1,g_1}},..., Sol^{m'_r,d'_r,m'_f,d'_f}_{BLC_{f_m,g_m}}) =$$
$$= Sol^{m_r+m'_r, d_r+d'_r, m_f+m'_f, d_f+d'_f}_{BLC_{f \circ (f_1,...,f_m), g \circ (g_1,...,g_m)}}$$

## 5. Fixed and Inertial Delays

**5.1 Definition** Let $f: B^m \to B$, $u \in S^{(m)}$, $x \in S$ and $d \geq 0$. The equation (see 4.3 b.2.3))
$$x(t) = f(u(t-d))$$
is called the *fixed limit condition* induced by $f$ ($FLC_f$). The limit condition defined by this equation is also called *pure*, *ideal* or *non-inertial*. A limit condition that is not pure is called *inertial*.

**5.2 Corollary** $FLC_f$ is deterministic, time invariant and constant. It satisfies
$$I^f_d \circ (I^{f_1}_{d'},..., I^{f_m}_{d'}) = I^{f \circ (f_1,...,f_m)}_{d+d'}, d \geq 0, d' \geq 0$$

## 6. Absolute Inertia

**6.1 Definition** The property
$$\overline{x(t-0)} \cdot x(t) \leq \bigcap_{\xi \in [t, t+\delta_r]} x(\xi) \tag{1}$$

$$x(t-0) \cdot \overline{x(t)} \leq \bigcap_{\xi \in [t, t+\delta_f]} \overline{x(\xi)} \quad (2)$$

true for $\delta_r \geq 0, \delta_f \geq 0$ is called the *absolute inertial condition* (*AIC*). We also call *AIC* the set $Sol_{AIC}^{\delta_r, \delta_f} \subset S$ defined by

$$Sol_{AIC}^{\delta_r, \delta_f} = \{x \mid x \text{ satisfies (1), (2)}\}$$

**6.2 Definition** A $LC_{f,g}$ $i$ with $\forall u, i(u) \subset Sol_{AIC}^{\delta_r, \delta_f}$ is called *absolute inertial limit condition* induced by $f, g$ ($AILC_{f,g}$).

**6.3 Theorem** Any $LC_{f,g}$ $i$ satisfying $\forall u, i(u) \wedge Sol_{AIC}^{\delta_r, \delta_f} \neq \emptyset$ defines the $AILC_{f,g}$ $i \wedge Sol_{AIC}^{\delta_r, \delta_f}$.

**6.4 Theorem** We consider $Sol_{BLC_{f,g}}^{m_r, d_r, m_f, d_f}$, the numbers $\delta_r \geq 0, \delta_f \geq 0$ and the statements:

a) $\forall u, Sol_{BLC_{f,g}}^{m_r, d_r, m_f, d_f}(u) \wedge Sol_{AIC}^{\delta_r, \delta_f} \neq \emptyset$

b) $d_r \geq d_f - m_f, d_f \geq d_r - m_r, \delta_r + \delta_f \leq m_r + m_f$

If $\max_{a \in B^m} f(a) \leq \min_{a \in B^m} g(a)$ then a) is true, otherwise a) and b) are equivalent.

**6.5 Definition** We suppose that $\forall u, Sol_{BLC_{f,g}}^{m_r, d_r, m_f, d_f}(u) \wedge Sol_{AIC}^{\delta_r, \delta_f} \neq \emptyset$. The function $Sol_{BLC_{f,g}}^{m_r, d_r, m_f, d_f} \wedge Sol_{AIC}^{\delta_r, \delta_f}$ is called *bounded absolute inertial limit condition* induced by $f, g$ ($BAILC_{f,g}$).

**6.6 Corollary** In the conditions from Theorems 4.8 and 6.4 the next formula is true:

$$(Sol_{BLC_{f,g}}^{m_r, d_r, m_f, d_f} \wedge Sol_{AIC}^{\delta_r, \delta_f}) \circ$$

$$(Sol_{BLC_{f_1, g_1}}^{m_r', d_r', m_f', d_f'} \wedge Sol_{AIC}^{\delta_r', \delta_f'}, ..., Sol_{BLC_{f_m, g_m}}^{m_r', d_r', m_f', d_f'} \wedge Sol_{AIC}^{\delta_r', \delta_f'}) =$$

$$= Sol_{BLC_{f \circ (f_1, ..., f_m), g \circ (g_1, ..., g_m)}}^{m_r + m_r', d_r + d_r', m_f + m_f', d_f + d_f'} \wedge Sol_{AIC}^{\delta_r, \delta_f}$$

**References**
[1] S. E. Vlad: *Towards a Mathematical Theory of the Delays of the Asynchronous Circuits*; Analele Universitatii din Oradea, Fascicola matematica, Tom IX (2002)
[2] S. E. Vlad: *Defining the Delays of the Asynchronous Circuits*; CAIM 2003, Oradea Romania (May 29-31, 2003)

Author's address: str Zimbrului, Nr.3, Bl. PB68, Et.2, Ap.11, 410430, Oradea, Romania